\documentclass[11pt,a4paper]{article}

\usepackage[utf8]{inputenc}
\usepackage[T1]{fontenc}
\usepackage[margin=2.5cm,top=3cm,bottom=2.8cm,headheight=15pt]{geometry}
\usepackage{xcolor}
\usepackage{graphicx}
\usepackage{hyperref}
\usepackage{fancyhdr}
\usepackage{titlesec}
\usepackage{amsmath,amssymb}
\IfFileExists{mathpazo.sty}{%
  \usepackage{mathpazo}
  \IfFileExists{newtxmath.sty}{%
    \usepackage[xcharter,bigdelims,vvarbb]{newtxmath}%
  }{}
  \IfFileExists{zlmtt.sty}{%
    \usepackage[scaled=1.1]{zlmtt}%
  }{%
  }
}{%
  \usepackage{lmodern}
}
\usepackage{booktabs}
\usepackage{caption}
\usepackage[numbers,sort&compress]{natbib}
\usepackage{tikz}
\usepackage{eso-pic}
\usepackage{colortbl}
\usepackage{float}
\usepackage{microtype}
\usepackage{multirow}
\usepackage{array}
\usepackage{makecell}
\usepackage{framed}
\usepackage{longtable}
\usepackage{hhline}
\usepackage{fancyvrb}
\usepackage{multicol}
\usepackage{tabularx}
\usepackage{subcaption}
\usepackage{cleveref}

\definecolor{MeshyLime}{HTML}{C5F955}
\definecolor{MeshyLimeDark}{HTML}{7DA82A}       
\definecolor{MeshyBlack}{HTML}{181818}
\definecolor{MeshyPink}{HTML}{FF3E8F}
\definecolor{MeshyDarkGray}{HTML}{2A2A2A}
\definecolor{MeshyMidGray}{HTML}{666666}
\definecolor{MeshyLightGray}{HTML}{F5F5F5}
\definecolor{MeshyAccentLine}{HTML}{B8E86E}     

\hypersetup{
    colorlinks=true,
    linkcolor=MeshyDarkGray,
    citecolor=MeshyLimeDark,
    urlcolor=MeshyPink,
    pdfauthor={Meshy AI},
    pdftitle={Meshy T2: Fast Native Mesh Generation with Flow Matching},
}



\RequirePackage{xspace}
\makeatletter
\DeclareRobustCommand\onedot{\futurelet\@let@token\@onedot}
\def\@onedot{\ifx\@let@token.\else.\null\fi\xspace}

 \def\vs{\emph{vs}\onedot}

\makeatother

\newcommand{\mymodel}{Meshy T2\xspace}
\newcommand{\methodname}{Meshy T2\xspace}

\renewcommand{\paragraph}[1]{\noindent\textbf{#1.}\hspace*{1em}}

\titleformat{\section}
  {\Large\bfseries\color{MeshyBlack}}
  {\color{MeshyLimeDark}\thesection}{0.6em}{}

\titleformat{\subsection}
  {\large\bfseries\color{MeshyBlack}}
  {\color{MeshyLimeDark}\thesubsection}{0.5em}{}

\titleformat{\subsubsection}
  {\normalsize\bfseries\color{MeshyDarkGray}}
  {\color{MeshyPink!70}\thesubsubsection}{0.5em}{}

\titlespacing*{\section}{0pt}{2.0ex plus 0.6ex}{1.0ex plus 0.3ex}
\titlespacing*{\subsection}{0pt}{1.5ex plus 0.4ex}{0.7ex plus 0.2ex}
\titlespacing*{\subsubsection}{0pt}{1.2ex plus 0.3ex}{0.5ex plus 0.1ex}

\fancypagestyle{titlepage}{%
  \fancyhf{}%
  \fancyfoot[C]{%
    {\color{MeshyLime}\rule{12pt}{1.5pt}}\enspace
    {\small\color{MeshyMidGray}\thepage}\enspace
    {\color{MeshyLime}\rule{12pt}{1.5pt}}%
  }%
}

\newcommand{\meshyaccentbar}{%
  \AddToShipoutPictureBG*{%
    \AtPageUpperLeft{%
      \raisebox{-3.5pt}{%
        \color{MeshyLime}\rule{\paperwidth}{3.5pt}%
      }%
    }%
  }%
}

\begin{document}

\meshyaccentbar
\thispagestyle{titlepage}
\vspace*{-20mm}

\noindent
\includegraphics[height=1.0cm]{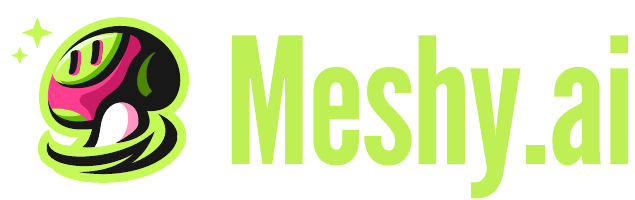}%

\noindent{\color{MeshyAccentLine}\rule{\textwidth}{1pt}}


\begin{center}
  {\fontsize{14}{18}\selectfont\bfseries\color{MeshyBlack}%
    Meshy T2: Fast Native Mesh Generation with Flow Matching
  }
\end{center}

\renewcommand{\thefootnote}{\fnsymbol{footnote}}
\begin{center}
  \renewcommand{\thefootnote}{\fnsymbol{footnote}}%
  {\small\color{MeshyMidGray}%
    Jiale Xu\footnote{Equal contribution.},
    Rendong Liang\footnotemark[1],
    Yuhao Long,
    Siyuan Shen,
    Zangyueyang Xian,
    Xiaofei Wu,
    Zeyi Xu\footnote{Contributed to this project during an internship at Meshy AI.},
    Yuanming Hu
  }\\
  \vspace{1mm}
  {\small\color{MeshyMidGray}%
    Meshy AI%
  }
\end{center}
\renewcommand{\thefootnote}{\arabic{footnote}}
\setcounter{footnote}{0}

\noindent\includegraphics[width=\textwidth]{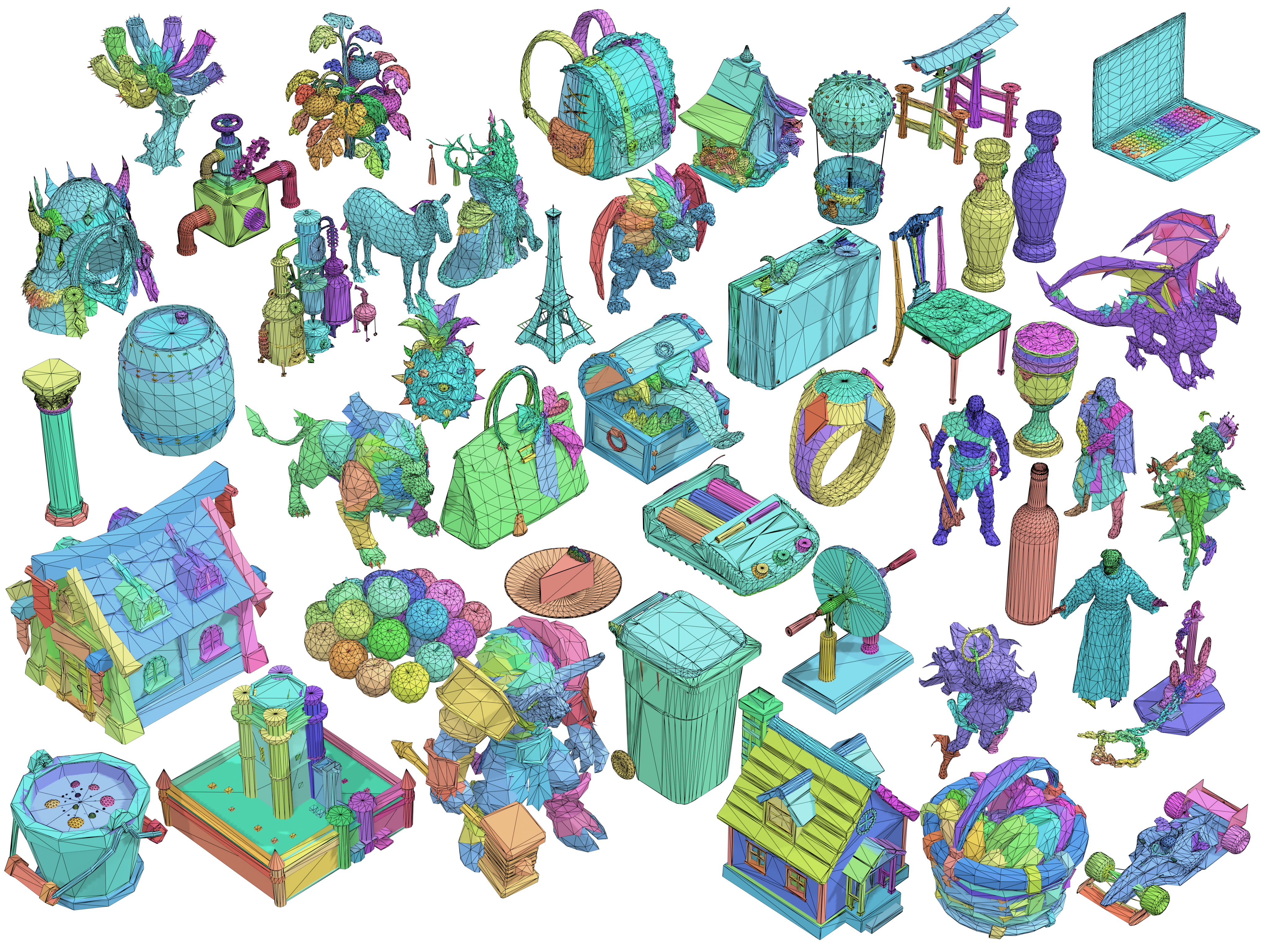}
\par\vspace{0.5mm}

\noindent
\begin{tikzpicture}
  \node[
    draw=MeshyAccentLine, fill=white,
    rounded corners=5pt, line width=0.8pt,
    inner sep=11pt,
    text width=\dimexpr\textwidth-30pt,
    align=justify,
  ] {%
    \noindent{\normalsize\bfseries Abstract}\par\vspace{0.2em}

    \footnotesize
    Polygonal meshes are the standard surface representation of modern 3D
    pipelines, and generating high-quality meshes with artist-style topology is
    essential for film, gaming, and interactive 3D applications. Mainstream
    approaches serialize a mesh into a token sequence and decode it
    autoregressively, which is slow at inference and sensitive to error
    accumulation, making them impractical for interactive asset creation. We
    present \methodname{}, a fast native mesh generation framework built on flow
    matching. At its core is a vertex-set mesh VAE that encodes a mesh into one
    continuous latent token per vertex and decodes vertices, edge connectivity,
    and face winding order in a single pass, preserving high-precision geometry
    and artist-authored topology without vertex quantization or welding.
    Generation proceeds as a coarse-to-fine cascade of two flow-matching models:
    an image-conditioned voxel flow first sketches the overall shape as a coarse
    occupancy scaffold, and a mesh flow then populates the scaffold with
    per-vertex latent tokens, conditioned on the image, the scaffold, and a
    requested vertex budget. This design delivers three practical capabilities:
    interactive generation speed through parallel flow-based synthesis;
    effective face-count control through the requested vertex budget; and native
    support for multi-part assets, whose components emerge directly from the
    generated connectivity. In our experiments, \methodname{} achieves
    state-of-the-art geometric fidelity and completes end-to-end image-to-mesh
    generation within a median of 6 seconds, over an order of magnitude faster
    than autoregressive baselines. Code and weights will be available at
    \url{https://github.com/meshy-dev/meshy-t2}.


  };
\end{tikzpicture}

\vspace{8mm}

\begin{figure*}[!t]
  \centering
  \includegraphics[width=\textwidth]{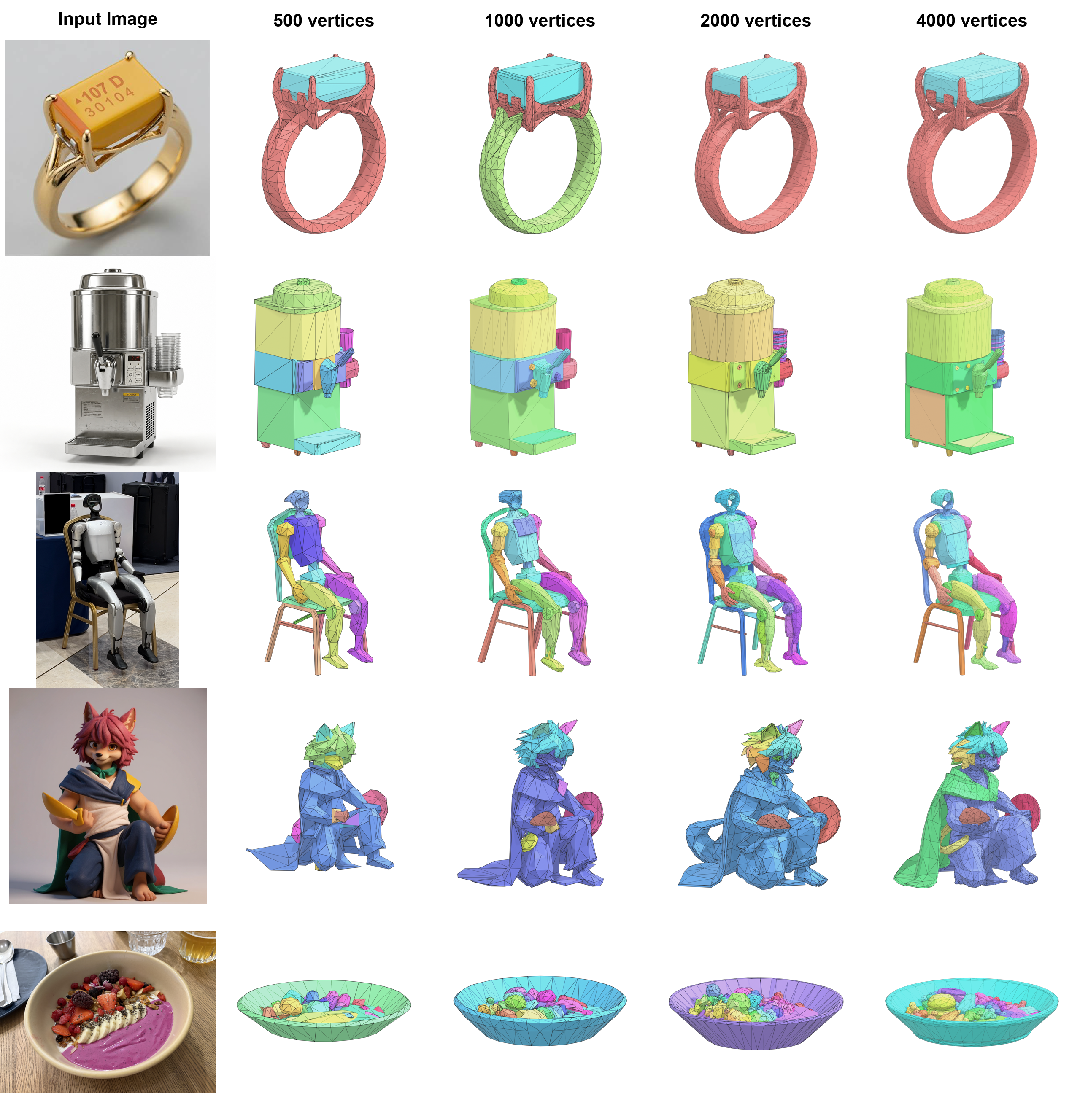}
  \caption{\textbf{\mymodel} generates compact meshes at interactive speed, with
  controllable face budgets and explicit connectivity. Because vertices and
  edges are generated jointly, multi-part assets naturally decompose into
  connected components without any separate part-wise generation.}
  \label{fig:teaser}
\end{figure*}

\newpage


\section{Introduction}
\label{sec:intro}

Polygonal meshes underpin virtually every real-time 3D application: game
engines, AR/VR systems, robotics simulators, and digital-content pipelines all
operate on vertices, edges, and faces for rendering, collision, editing,
storage, and transmission. Not every mesh serves these applications equally
well, however. A production-ready asset must capture the intended shape with
as few primitives as possible while preserving sharp edges, thin structures,
and semantically meaningful part boundaries: redundant faces inflate
bandwidth, memory footprint, rasterization cost, and editing effort, whereas
overly aggressive simplification sacrifices geometric fidelity. Meshes of
this quality are still predominantly authored by hand, with skilled artists
modeling or retopologizing each asset primitive by primitive---a slow and
costly workflow that cannot keep pace with the growing demand for 3D content.
Automatically generating compact, valid, and ready-to-use meshes is therefore
a central challenge in scalable 3D content creation.

Most recent high-quality 3D generative systems do not generate such meshes
directly. Instead, they first learn an implicit or volumetric geometry
representation and then extract an explicit surface as a post-processing
step~\cite{hong2023lrm,li2023instant3d,xu2024instantmesh,xu2024freesplatter}.
Representative methods such as 3DShape2VecSet~\cite{zhang20233dshape2vecset},
TRELLIS~\cite{xiang2025structured}, and LATTICE~\cite{xiang2025native} model
geometry as neural fields decoded from vector-set or sparse-voxel latents, and
obtain meshes through iso-surface extraction, typically Marching
Cubes~\cite{lorensen1987marching} and its
variants~\cite{schaefer2005dual,shen2023flexicubes,luo2025faithful}. This
design is attractive because implicit fields are continuous and easy to
optimize with image or 3D supervision. The extracted meshes, however, are far
from ready to use. Since extraction tessellates the surface according to grid
resolution rather than geometric structure, the resulting meshes typically
carry hundreds of thousands of near-uniform triangles---far too dense for
rendering, editing, and animation pipelines. Simplification algorithms can
reduce the face count, but as purely geometric post-processing they produce
irregular triangulations that poorly respect sharp features and part
structure, still well short of the clean, artist-style topology that
production assets demand.

A second line of work therefore generates meshes directly, treating mesh
generation as a sequence modeling problem. MeshGPT~\cite{siddiqui2023meshgpt}
pioneered this paradigm: it serializes faces into discrete coordinate
sequences and models them with an autoregressive transformer, directly
producing compact, artist-like topology that preserves sharp features far
better than iso-surface extraction. Since the resulting token streams are
extremely long, follow-up works have largely focused on compressing the
serialization---EdgeRunner~\cite{tang2025edgerunner},
TreeMeshGPT~\cite{lionar2025treemeshgpt}, and
Mesh-Silksong~\cite{songtopology} design topology-aware traversals that
maximize edge reuse, while BPT~\cite{weng2024bpt} and
DeepMesh~\cite{zhao2025deepmesh} compress coordinates through block-wise
indexing---and on scaling the paradigm with shape conditioning and larger
architectures~\cite{chen2024meshanything,chen2024meshanythingv2,chen2024meshxl,hao2024meshtron,xu2026meshweaver}.
Yet better compression does not resolve the deeper mismatch: a mesh is an
unordered whole---vertices and faces coupled by adjacency, with no canonical
linear order---and flattening it into a 1D token sequence is fundamentally
unnatural. As a consequence, inference remains inherently sequential and
grows expensive with the face budget, and mesh validity becomes an emergent
property of a long sampled sequence: a local sampling mistake can propagate
into incomplete surfaces, inconsistent face winding, or cracks between
adjacent regions.

Diffusion- and flow-based formulations are a natural response to this
mismatch: by generating all primitives in parallel instead of imposing a
traversal order, they respect the unordered structure of a mesh, decouple
inference cost from mesh size, and avoid sequential error accumulation.
Existing methods differ mainly in the mesh latent representation they
denoise. MeshCraft~\cite{he2025meshcraft} and
MeshFlow~\cite{sun2026meshflowequivariant} operate on face-level tokens,
which duplicate shared vertices across adjacent faces and leave cross-face
consistency to emerge from sampling. TriFlow~\cite{li2026triflow},
LATO~\cite{zhao2026lato}, and LATO.2~\cite{long2026lato2} encode meshes into
sparse-voxel latents, re-deriving mesh primitives from volumetric features at
decoding time. MeshFlow~\cite{li2026meshflowvae} and Nexus~\cite{wang2026nexus} adopt
per-vertex tokens---the granularity closest to the mesh itself---though
Nexus still generates geometry and topology in decoupled stages.
Nevertheless, two gaps remain. First, the mesh representations are seldom
designed for exactness: vertex quantization and heuristic face
reconstruction---e.g., recovering triangles as 3-cliques of a predicted edge
graph---introduce artifacts, so the detail and topology of the original mesh
cannot be preserved faithfully through the latent space, and the decoded
asset often requires post-processing to repair invalid connections,
overlapping faces, or non-manifold edges before use. Second, most of
these methods condition on point clouds sampled from an existing surface and
thus effectively perform retopology; the additional challenges of
image-to-mesh generation, where no dense geometry is available at inference
time, have not been systematically addressed.

We introduce \methodname{}, a flow-based framework that closes both gaps:
it builds on a carefully designed, nearly lossless vertex-set mesh
representation extending SpaceMesh~\cite{shen2024spacemesh}, and it targets
image-to-mesh generation directly, producing compact, ready-to-use meshes at
interactive speed. Given
a reference image, our method first predicts a coarse voxel scaffold that
captures the global object layout, and then, conditioned on both the image
and the scaffold, generates the vertices jointly with their
edge connectivity and face winding order. This design yields three practical
capabilities. First, generation is parallel over latent vertex tokens,
completing end-to-end image-to-mesh generation within ten seconds. Second,
the user sets the
vertex budget before decoding, which directly controls the complexity of the
generated mesh without post-hoc simplification. Finally, the mesh VAE is
nearly lossless: vertex positions remain continuous, coincident vertices are
never welded, and the artist-authored connectivity is preserved exactly.
Generated meshes therefore retain high-precision geometric detail and
faithful artist-style topology, and naturally decompose into connected
components, so multi-part assets are produced directly without a separate
component-wise generation or stitching stage.

The contributions of this report are summarized as follows:
\begin{itemize}
    \item We design a nearly lossless vertex-set mesh VAE extending
    SpaceMesh: coordinates are never quantized, coincident vertices are never
    welded, and vertices, edge connectivity, and face winding order are
    recovered jointly in a single decoding pass, faithfully preserving the
    geometry, artist-authored topology, and part structure of source meshes.
    \item We build a two-stage flow-matching pipeline for direct
    image-to-mesh generation: an image-conditioned voxel flow sketches a
    coarse occupancy scaffold, and a mesh flow populates it with per-vertex
    latent tokens under optimal-transport-assigned positional encodings,
    completing end-to-end generation within ten seconds.
    \item We provide strong face-count control by sampling a requested number
    of vertex slots before decoding, which directly determines the expected
    face budget without post-hoc simplification.
    \item Experiments show that \methodname{} achieves state-of-the-art
    geometric fidelity on both retopology and image-to-mesh tasks while
    running over an order of magnitude faster than autoregressive baselines.
\end{itemize}










\section{Method}
\label{sec:method}

\methodname{} generates meshes in a latent space where every token
corresponds to exactly one vertex. This representation is established by a
vertex-set mesh VAE, whose encoder maps an explicit mesh into a per-vertex
latent set and whose decoder recovers vertices, edges, and oriented faces
from such a set in a single pass (Sec.~\ref{sec:model:mesh_vae}).
Generation itself is a coarse-to-fine cascade of two flow-matching
models~\cite{lipman2022flow}, instantiated with the linear interpolation
schedule of Rectified Flow~\cite{liu2022flow}. Given a reference image, a voxel flow first
sketches the overall shape as a $64^3$ occupancy scaffold
(Sec.~\ref{sec:model:voxel_flow}); a latent flow then populates this scaffold
with per-vertex latent tokens, guided jointly by the image, the voxel scaffold, and
the requested vertex budget (Sec.~\ref{sec:model:mesh_flow}). Decoding the
generated latent set with the VAE decoder yields the final mesh.

\subsection{Mesh VAE}
\label{sec:model:mesh_vae}

\newcommand{\pipelinevoxelencode}{%
  \includegraphics[width=\textwidth,page=1,trim=142 370 183 0,clip]{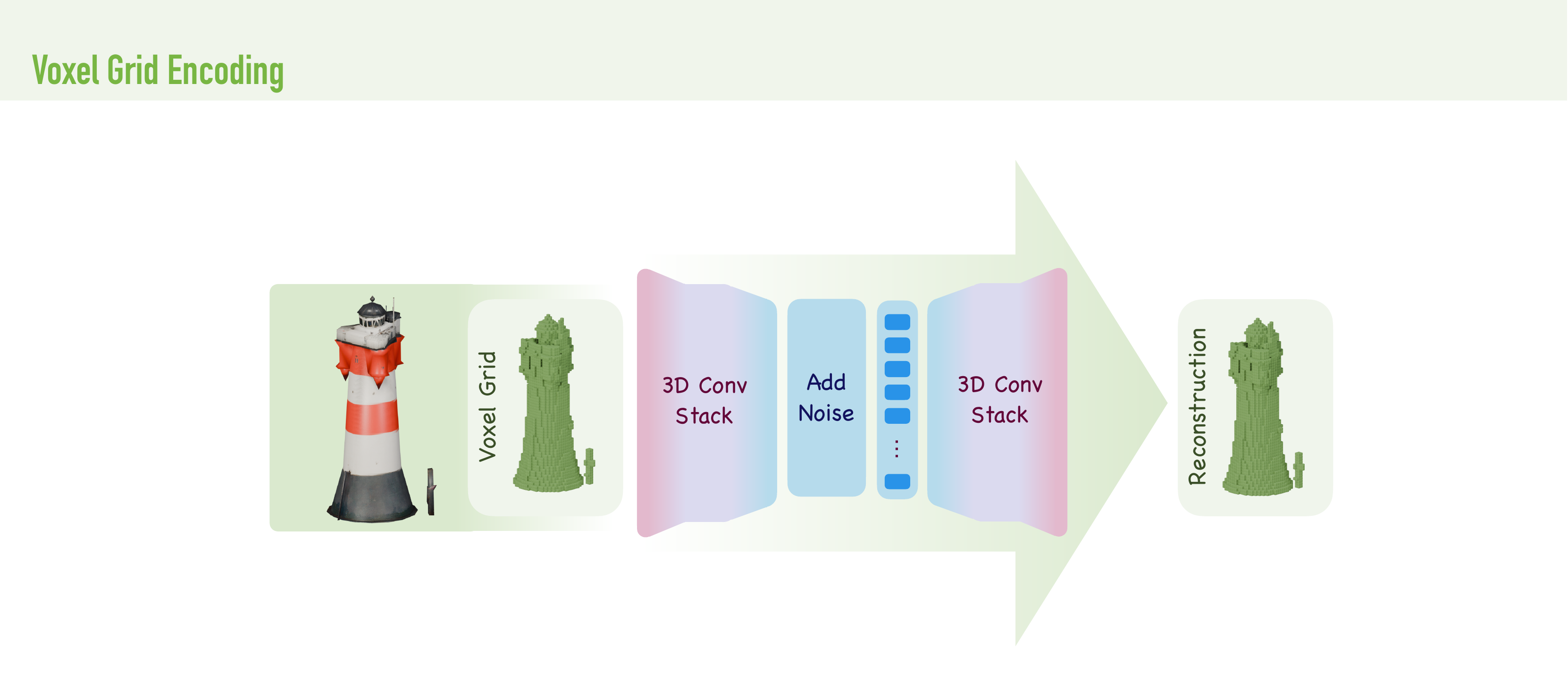}}
\newcommand{\pipelinemeshencode}{%
  \includegraphics[width=\textwidth,page=2,trim=142 182 183 0,clip]{figures/pipeline.pdf}}
\newcommand{\pipelinevoxelgenerate}{%
  \includegraphics[width=\textwidth,page=3,trim=142 209 183 0,clip]{figures/pipeline.pdf}}
\newcommand{\pipelinemeshgenerate}{%
  \includegraphics[width=\textwidth,page=4,trim=142 140 183 0,clip]{figures/pipeline.pdf}}

\begin{figure*}[!t]
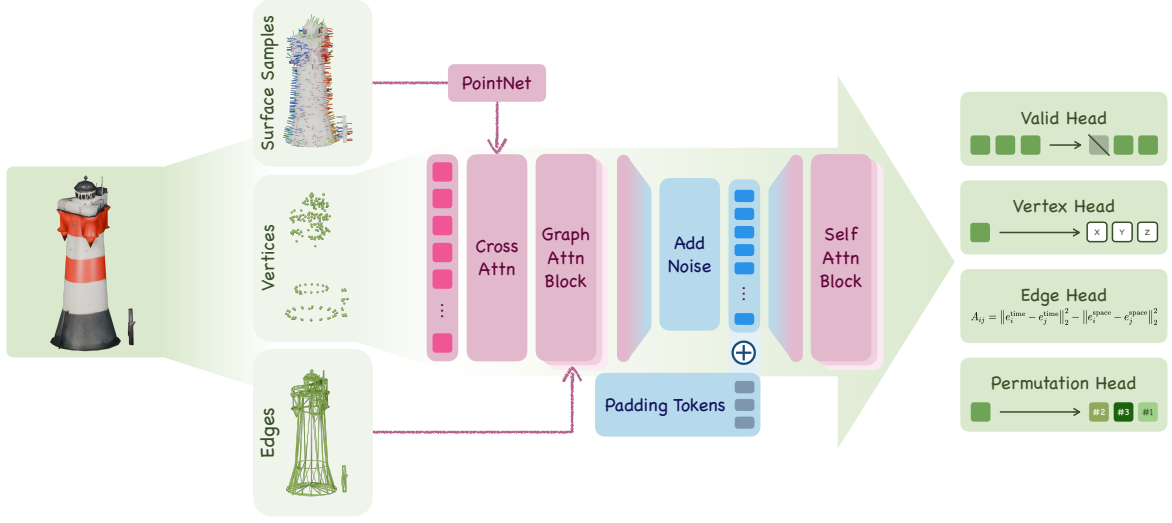

\centering
\vspace{-0.4em}
\pipelinemeshencode
\vspace{-0.6em}
\caption{\textbf{Vertex-set mesh autoencoder training framework.}
The encoder builds one latent token per ground-truth vertex by combining
surface-sample voxel context with vertex queries and graph attention over the
true mesh edges. The decoder reconstructs vertex coordinates, undirected
edges, and local halfedge permutations from the latent tokens. The final
mesh is assembled from the predicted edges and halfedge successor
mappings.}
\label{fig:vae_training_framework}
\end{figure*}

Given a triangle mesh $\mathcal{M} = (\mathcal{V}, \mathcal{F})$ with
vertices $\mathcal{V} = \{x_i\}_{i=1}^{V}$, $x_i \in \mathbb{R}^3$, and
triangle faces $\mathcal{F}$, the VAE encodes $\mathcal{M}$ into a latent set
$\mathcal{Z} = \{z_i\}_{i=1}^{V}$ with exactly one token
$z_i \in \mathbb{R}^{C}$ per vertex, and decodes both the vertex positions
and the topology from $\mathcal{Z}$ alone
(Fig.~\ref{fig:vae_training_framework}). Most prior mesh generators
quantize vertex coordinates onto a discrete grid and weld vertices that
coincide after quantization, which blurs fine geometric detail, silently
rewires the original connectivity, and fuses parts that merely touch. Our
VAE does neither: positions are regressed continuously and coincident
vertices keep distinct tokens, so the high-precision geometry, the
artist-authored topology, and the part structure of the source mesh are
preserved exactly. Following
SpaceMesh~\cite{shen2024spacemesh}, we represent the topology at two levels:
the undirected edge set $\mathcal{E}$, predicted through a spacetime
embedding of adjacency, and the oriented faces, encoded as per-vertex
halfedge successor permutations.

\paragraph{Encoder}
The encoder produces one latent token per ground-truth vertex from two
inputs: a sparse voxel context that summarizes the surface, and one query
token per vertex. The context is produced by a local
PointNet~\cite{qi2017pointnet}
that pools surface samples---positions and normals, drawn preferentially
along mesh edges---into features on the occupied cells of a $256^3$ sparse
voxel grid; in our experiments, this point context substantially accelerates
the convergence of the VAE. Each vertex query is initialized from Fourier
features~\cite{tancik2020fourier} of its continuous position, and all
encoder attention applies 3D
RoPE~\cite{su2024roformer}. The queries are refined by one
cross-attention into the voxel context, followed by a stack of
interleaved graph-attention and self-attention layers: the graph attention
performs message passing restricted to the ground-truth edge set
$\mathcal{E}$, while the self-attention operates over the full vertex set. A
final projection maps each vertex feature to its latent $z_i$.

\paragraph{Decoder}
The decoder is a pure set decoder: it consumes latent tokens without any
positional encoding and, for every vertex, predicts a continuous position
$\hat{x}_i$ together with an edge embedding $e_i$ and a face embedding
$f_i$. A shared self-attention trunk processes the latent set
$\mathcal{Z}$, followed by two separate self-attention branches: a vertex
branch, whose head regresses the position $\hat{x}_i$, and a topology
branch, whose features are mapped into the edge and face embeddings. How
these embeddings are turned into edges and faces is described next.

\paragraph{Edge prediction}
The edge head realizes the spacetime view of adjacency from
SpaceMesh~\cite{shen2024spacemesh}. The edge embedding is split into a
spatial half and a temporal half,
$e_i = (e_i^{\mathrm{space}}, e_i^{\mathrm{time}})
\in \mathbb{R}^{d_e} \times \mathbb{R}^{d_e}$, and every
vertex pair is scored by the Minkowski-style logit
\begin{equation}
    A_{ij} \;=\;
    \bigl\| e_i^{\mathrm{time}} - e_j^{\mathrm{time}} \bigr\|_2^2
    \;-\;
    \bigl\| e_i^{\mathrm{space}} - e_j^{\mathrm{space}} \bigr\|_2^2 ,
    \label{eq:edge_logit}
\end{equation}
so that an edge is predicted between $i$ and $j$ exactly when their temporal
separation exceeds their spatial separation. The adjacency matrix
$A \in \mathbb{R}^{V \times V}$ is symmetric by construction, and only the
strict upper triangle is supervised against the ground-truth edge set
$\mathcal{E}$ with a class-balanced binary cross-entropy,
\begin{equation}
    \mathcal{L}_{\mathrm{edge}} \;=\;
    \frac{1}{Z}
    \Bigl[\,
    \sum_{\{i,j\} \in \mathcal{E}} \operatorname{softplus}(-A_{ij})
    \;+\;
    \lambda \!\!\sum_{\substack{i<j \\ \{i,j\} \notin \mathcal{E}}}\!\!
    \operatorname{softplus}(A_{ij})
    \,\Bigr],
    \label{eq:edge_loss}
\end{equation}
where $\lambda$ down-weights the abundant negative pairs to balance the two
classes and $Z$ is the corresponding effective sample count. The loss is
macro-averaged over the meshes of a batch and evaluated without dense
$V \times V$ supervision targets, which would not fit in GPU memory for
high-vertex-count meshes.

\paragraph{Face prediction}
Following SpaceMesh~\cite{shen2024spacemesh}, the face head assembles
oriented faces by predicting, for every vertex $i$, the cyclic order of its
triangle fan, i.e., which neighbor follows which when walking around $i$. Each
oriented triangle that touches $i$ can be written
as $(p, i, n)$, where $p$ and $n$ are the two neighbors it connects to $i$,
and its orientation says that, walking around $i$, the edge to $p$ is
immediately followed by the edge to $n$. Collecting this relation over all
triangles at $i$ defines a successor mapping $\pi_i$ on the neighbors
$\mathcal{N}(i)$,
\begin{equation}
    \pi_i(p) \;=\; n
    \qquad \text{for every triangle } (p, i, n) \in \mathcal{F}.
    \label{eq:face_target}
\end{equation}
Read together, $\mathcal{E}$ and $\{\pi_i\}$ losslessly encode the oriented
face set---every triangle appears as three mutually consistent successor
links---so face prediction reduces to predicting one successor mapping per
vertex. This representation is well defined only for manifold meshes, where
each neighbor has a unique successor; non-manifold meshes in the dataset are
repaired in advance by splitting the offending edges and vertices.

SpaceMesh assumes watertight meshes: every fan is closed, and $\pi_i$ is a
cyclic permutation of $\mathcal{N}(i)$. Artist-created assets, however,
frequently contain open boundaries. We therefore extend the domain of
$\pi_i$ with a NULL element $\varnothing$. At a boundary vertex, whose fan
is an open strip rather than a closed loop, the last edge maps to
$\varnothing$ and $\varnothing$ maps back to the first edge; at an interior
vertex, $\varnothing$ simply maps to itself. Every fan thus closes into a
proper permutation on $\mathcal{N}(i) \cup \{\varnothing\}$, and open,
non-watertight surfaces are represented exactly rather than approximated or
discarded.

The face head predicts a soft version $P_i$ of $\pi_i$ from the face
embeddings. The embedding of vertex $i$ consists of three
$d_f$-dimensional vectors,
$f_i = (f_i^{\mathrm{root}}, f_i^{\mathrm{prev}}, f_i^{\mathrm{next}})$,
which let the vertex act as the center of a fan, as a predecessor, and as a
successor; two learnable vectors
$f_{\varnothing}^{\mathrm{prev}}, f_{\varnothing}^{\mathrm{next}}$ play the
latter two roles for the NULL element. For the fan of vertex $i$, every
ordered pair $p, q \in \mathcal{N}(i) \cup \{\varnothing\}$ is scored
by
\begin{equation}
    \Phi_i[p,q] \;=\;
    \mathbf{1}^{\top}
    \bigl(
    f^{\mathrm{root}}_{i} \odot
    f^{\mathrm{prev}}_{p} \odot
    f^{\mathrm{next}}_{q}
    \bigr),
    \label{eq:face_score}
\end{equation}
the element-wise product of the three role vectors summed over channels,
and Sinkhorn iterations~\cite{cuturi2013sinkhorn} in log space normalize $\Phi_i$ into an
approximately doubly stochastic matrix $P_i$---a differentiable relaxation
of a permutation.

For supervision, the target mappings $\pi_i$ are extracted once per
training mesh after propagating a consistent orientation across adjacent
faces. The face head is trained with the negative log-likelihood
\begin{equation}
    \mathcal{L}_{\mathrm{face}} \;=\;
    -\,\frac{1}{\sum_i (D_i + 1)}
    \sum_{i} \;\sum_{p \,\in\, \mathcal{N}(i) \cup \{\varnothing\}}
    \log P_i\bigl[p,\, \pi_i(p)\bigr],
    \label{eq:face_loss}
\end{equation}
where $D_i = |\mathcal{N}(i)|$ is the degree of vertex $i$. The loss covers
the triangle-induced successor links and the NULL transitions alike, and its
evaluation is grouped by vertex degree, so no dense
$[V, D_{\max}, D_{\max}]$ tensor is ever built.

\paragraph{Mesh assembly}
At inference time, the decoder outputs are converted into an explicit mesh
in three steps. First, the predicted positions $\hat{x}_i$ become the
vertices, and every pair with $A_{ij} > 0$ becomes an undirected edge.
Second, each soft matrix $P_i$ is rounded into a hard successor mapping
$\hat{\pi}_i$ by a linear assignment constrained to form a single fan rather
than several disjoint sub-cycles. Third, oriented triangles are read off from the hard
successor mappings, with each directed halfedge used at most once so that
the resulting faces are consistently oriented. The output is a complete
mesh---vertices, edges, and oriented faces---produced in a single
decoding pass. Since connectivity is predicted explicitly, connected
components come for free: a multi-part asset is decoded as a single mesh
whose parts are already separated in the vertex--edge graph, without any
component-wise generation or stitching.

\paragraph{Training objective}
The autoencoder is optimized end to end with a weighted sum of the vertex,
edge, and face losses,
\begin{equation}
    \mathcal{L}_{\mathrm{VAE}} =
    w_v \mathcal{L}_{\mathrm{vertex}} +
    w_e \mathcal{L}_{\mathrm{edge}} +
    w_f \mathcal{L}_{\mathrm{face}} ,
    \label{eq:vae_loss}
\end{equation}
where $\mathcal{L}_{\mathrm{vertex}}$ is the MSE between the predicted
positions $\hat{x}_i$ and the ground truth $x_i$, and
$\mathcal{L}_{\mathrm{edge}}$ and $\mathcal{L}_{\mathrm{face}}$ are given in
Eqs.~\eqref{eq:edge_loss} and~\eqref{eq:face_loss}. Concrete loss weights and
architectural dimensions are given in the implementation details at the end
of this section.

\paragraph{Implementation details}
All attention layers use a width of 1024 with 16 heads; the encoder consumes
204{,}800 surface samples per mesh and stacks six pairs of interleaved
graph-attention and self-attention layers, and the decoder uses a shared
trunk of 12 blocks followed by vertex and topology branches of 4 blocks
each. The latent has $C = 32$ channels, the edge and face embedding sizes
are $d_e = 16$ and $d_f = 16$, and $P_i$ is normalized with 10 Sinkhorn
iterations. Training uses loss weights $w_v = 100$ and $w_e = w_f = 10$ together
with scale and rotation augmentation.

For the point encoder, standard scatter-based mean-pooling into sparse voxels
relies on atomic accumulation and repeated index gathering, which severely slows
down training speed. To eliminate this bottleneck, we sort points once per
forward pass by their (batch, voxel) keys into a contiguous voxel-major layout
with CSR segment offsets. All subsequent poolings in both forward and backward
passes then execute contiguous segment reductions and broadcasting without
atomic operations.

\subsection{Stage I: Image-Conditioned Voxel Flow}
\label{sec:model:voxel_flow}

Our goal is image-to-mesh generation. A natural first attempt is to train a
single flow that generates the per-vertex latent set directly from the
image; in our experiments, however, this performed poorly. The difficulty is
not surprising: a single image constrains the global 3D shape only weakly, while
the latent set is a large, unordered collection of fine-grained tokens, so
one model has to resolve the global layout, the vertex placement, and the
local tessellation all at once from an ambiguous 2D observation. We
therefore split generation into two stages. The first stage, described in
this section, synthesizes only a coarse geometric scaffold: a binary
occupancy grid $\mathbf{O} \in \{0,1\}^{64 \times 64 \times 64}$ that fixes
where the object occupies space, but says nothing about its tessellation.
The second stage (Sec.~\ref{sec:model:mesh_flow}) then generates the
per-vertex latent set anchored on this scaffold. With this split, the
scaffold model concentrates on global shape and image alignment, while the
latent-set model concentrates on tessellation and topology.

Following the sparse-structure generation strategy of
TRELLIS~\cite{xiang2025structured}, we do not model the binary occupancy
grid directly. Generation instead runs in the continuous latent space of a
pretrained Voxel VAE, which removes much of the spatial redundancy of the
raw grid and provides a smooth distribution for flow-based modeling.

\begin{figure*}[!t]
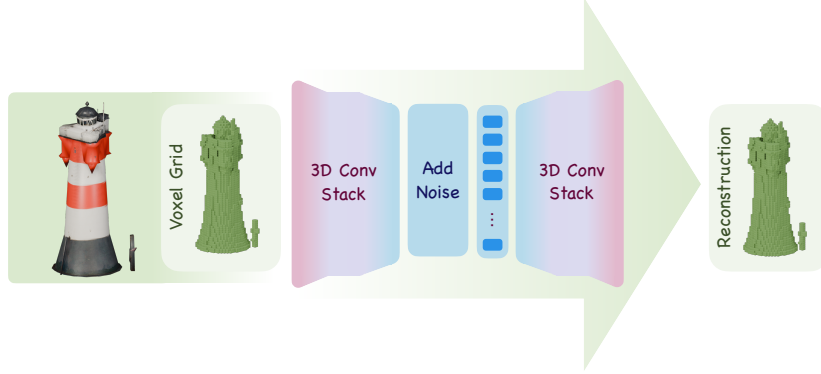

\centering
\vspace{-0.4em}
\pipelinevoxelencode
\vspace{-0.6em}
\caption{\textbf{Voxel-grid encoding.}
The Voxel VAE is trained on $64^3$ occupancy grids voxelized directly from
artist-authored meshes. Two stride-2 convolutional stages parameterize an
$8$-channel spatial Gaussian posterior at resolution $16^3$; the mirrored
decoder reconstructs occupancy logits through two 3D pixel-shuffle upsampling
stages, supervised by binary cross-entropy with a small KL penalty.}
\label{fig:voxel_vae_architecture}
\end{figure*}

\paragraph{Voxel VAE}
The Voxel VAE is a dense 3D convolutional VAE
(Fig.~\ref{fig:voxel_vae_architecture}). The encoder compresses the
occupancy grid through two stride-2 stages into a spatially factorized
Gaussian posterior with an 8-channel mean and log-variance at resolution
$16^3$; the mirrored decoder maps a latent grid back to occupancy logits
through two 3D pixel-shuffle upsampling stages, and the binary scaffold is
recovered by thresholding the predicted occupancy probability at $0.5$.
The VAE is trained with a binary cross-entropy reconstruction loss on the
logits and a small KL penalty. The training grids are voxelized directly from the
original artist-authored meshes, without remeshing or procedural
retopology. Once trained, the VAE is frozen, and every occupancy grid is
represented by its posterior mean
$\mathbf{z}_{\mathrm{voxel}} = \boldsymbol{\mu}_{\phi}(\mathbf{O})
\in \mathbb{R}^{8 \times 16^3}$---a deterministic encoding that retains the
regularized latent geometry.

\begin{figure*}[!t]
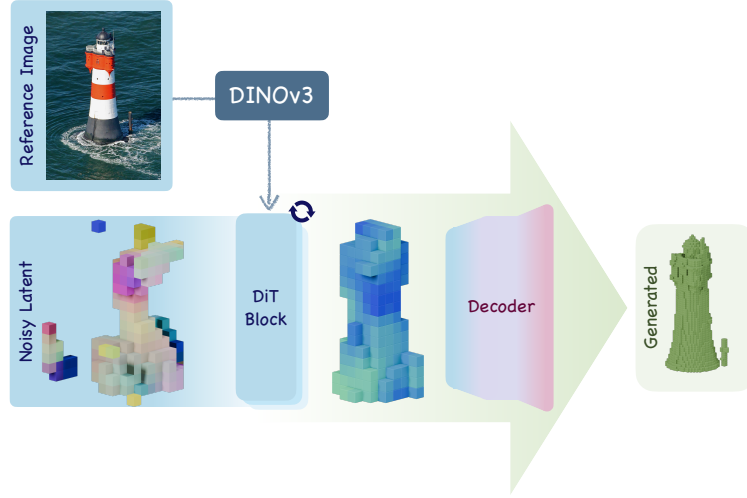

\centering
\vspace{-0.4em}
\pipelinevoxelgenerate
\vspace{-0.6em}
\caption{\textbf{Voxel-grid generation.}
With the Voxel VAE frozen, its standardized posterior mean is flattened into
$16^3=4096$ spatial tokens as the clean endpoint $\mathbf{x}_0$, and a
time-modulated Transformer velocity field is trained with flow matching under
DINOv3 image conditioning. At inference, the generated latent grid is decoded
by the frozen VAE and thresholded into the binary occupancy scaffold that
conditions the second stage.}
\label{fig:voxel_flow_generation}
\end{figure*}

\paragraph{Flow model}
The scaffold generator is a Transformer velocity field $f_{\theta}$ over
the latent grid (Fig.~\ref{fig:voxel_flow_generation}). The posterior mean is
standardized with fixed channel-wise
statistics and flattened into $16^3 = 4096$ tokens, each keeping its
three-dimensional coordinate; this sequence is the clean endpoint
$\mathbf{x}_0$ of the flow. The noisy tokens are processed by a stack of
time-modulated Transformer blocks: self-attention with 3D positional
encoding models dependencies among the spatial cells, the reference
image---encoded by a frozen DINOv3 backbone~\cite{simeoni2025dinov3}---is
injected into every block through cross-attention with voxel tokens as
queries and image features as keys and values, and the timestep is
injected through AdaLN modulation~\cite{peebles2023scalable}. A final
projection yields one
velocity prediction per cell,
$\widehat{\mathbf{v}} = f_{\theta}(\mathbf{x}_t, t,
\mathbf{c}_{\mathrm{img}})$, where $\mathbf{c}_{\mathrm{img}}$ denotes the
encoded image condition. The model is trained with velocity-prediction
flow matching. At inference, the
generated latent grid is decoded by the frozen Voxel VAE and thresholded
into the binary occupancy scaffold that conditions the second stage.

\paragraph{Implementation details}
The VAE encoder and decoder use two residual blocks at each resolution
level, with channel-wise normalization, SiLU activations, and
zero-initialized second convolutions so that each block starts near an
identity mapping; the KL weight is $\lambda_{\mathrm{KL}} = 10^{-4}$. The
flow Transformer has a hidden dimension of 1536 with 28 blocks and 12
attention heads, and each block combines 3D positional self-attention,
cross-attention to the image features, timestep-conditioned modulation, and
a SwiGLU feed-forward network~\cite{shazeer2020glu}.

\subsection{Stage II: Image-and-Voxel-Conditioned Mesh Flow}
\label{sec:model:mesh_flow}

\begin{figure*}[!t]
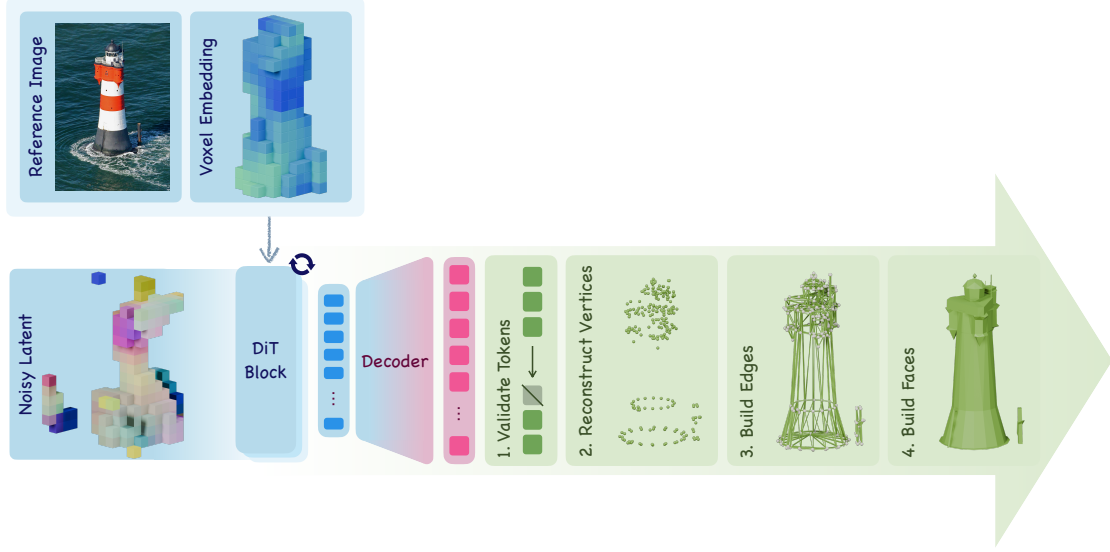

\centering
\vspace{-0.4em}
\pipelinemeshgenerate
\vspace{-0.6em}
\caption{\textbf{Mesh generation.}
Frozen VAE latents provide the clean target set $\mathbf{x}_0$. Flow matching
constructs $\mathbf{x}_t$ from $\mathbf{x}_0$, Gaussian noise $\mathbf{x}_1$,
and timestep $t$, then predicts velocity using a single-stream DiT over latent,
image, and voxel tokens. Sobol OT assigns spatial PE to unordered latent tokens,
while image, voxel, timestep, and count conditions control generation and
primitive budget.}
\label{fig:flow_training_framework}
\end{figure*}

The second generative stage operates on the latent sets produced by our
Mesh VAE (Fig.~\ref{fig:flow_training_framework}). Let
$\mathbf{x}_0=\{z_i\}_{i=1}^{N}$ denote the clean latent set
of a mesh, one $C$-channel token per vertex. We train a flow-matching model to
map Gaussian noise to $\mathbf{x}_0$ under image, voxel, and vertex count conditions.

\paragraph{Image and voxel condition}
Image conditioning uses a frozen DINOv3 image
encoder~\cite{simeoni2025dinov3} at $768 \times 768$ resolution. With a patch
size of 16, each image contributes a $48 \times 48$ grid, or 2304 image
tokens, which a trainable projection maps to the flow hidden size.
Voxel conditioning starts from the $64^3$ binary occupancy grid: during
training it is voxelized from the ground-truth mesh, while at inference it
is the scaffold generated by the first stage. The grid is encoded by the
Voxel VAE encoder (Sec.~\ref{sec:model:voxel_flow}) into a $16^3$ latent
grid, yielding 4096 voxel condition tokens per sample after projection to
the flow hidden size. The voxel encoder is not frozen but fine-tuned at a
$0.25\times$ learning rate.

\paragraph{Vertex count condition}
Face-count control is a practical necessity: real-time and mobile
applications impose hard polygon budgets on their assets, and
level-of-detail pipelines require the same shape at several prescribed
complexities. Since our latent tokens are vertex-level, we can control the face count
indirectly by conditioning generation on the vertex count: for a closed
triangle mesh, Euler's formula combined with the fact that every edge is
shared by exactly two triangles yields
$F = 2V - 4 \approx 2V$~\cite{botsch2010polygon}. We
represent the requested latent slot count $N$ as a Fourier embedding of
$N/N_{\max} \in [0, 1]$, where $N_{\max}$ is a preset maximum vertex count, project it
to the time-embedding dimension, and add it to the
time embedding. Exact-count control, however, is difficult to learn: the
vertex counts covered by our dataset are limited, so a model asked to spend
exactly $N$ vertices on an arbitrary shape has no reliable way to comply.
We therefore relax exact-count control into range control. During training,
zero-valued pad tokens are randomly appended to the latent set up to a
ratio $p$, and every token gains an additional existence channel set to
$+1$ for real tokens and $-1$ for pads, so the flow state has $C{+}1$
channels; the count condition is set to the padded total token count. Pad
tokens are supervised like real tokens, with the all-zero latent and a
$-1$ existence channel as their clean target, which is precisely how the
model learns to switch off latent slots it does not need. At inference,
only the generated tokens with a positive existence channel are kept as
real vertices and passed to the VAE decoder, while the rest are discarded
as pads. Conditioned on a budget of $N$ latent slots, the model thus
generates a mesh whose vertex count falls within $[N/(1{+}p),\, N]$.

\paragraph{Flow model}
The flow model is a single-stream DiT~\cite{peebles2023scalable}: the
latent, image, and voxel tokens are concatenated into one sequence and
processed jointly by self-attention, and only the latent-token outputs are
gathered for the flow prediction. All tokens share a unified 3D RoPE, whose
positions are assigned as described in the next paragraph. The time
embedding and the vertex count embedding are injected through AdaLN
modulation. The model is trained with
velocity-prediction flow matching, with one timestep per mesh drawn from a
logit-normal distribution~\cite{esser2024scaling}. During training, the
image, voxel, and count
conditions are independently dropped with probabilities 0.2, 0.3, and 0.2,
and all conditions are jointly dropped with an additional probability of
0.05, forming the partially and fully unconditional branches needed for
classifier-free guidance~\cite{ho2022classifier}. A dropped image or voxel
condition is omitted from the sequence entirely rather than replaced by
zeroed tokens, while the additive count condition is zeroed.

\paragraph{Latent PE via Optimal Transport}
\label{sec:method:pe}
Image and voxel tokens carry natural spatial coordinates and share a
unified 3D RoPE, with the image patches placed on a separate slice of the
coordinate volume to avoid overlap. The latent tokens, however, are an
unordered set with no native positions, and leaving them unencoded is
harmful: every latent slot starts from i.i.d.\ Gaussian noise, so the
slots are mutually indistinguishable, and any permutation of the vertex
latents is an equally valid generation target under the same condition.
The velocity field is then forced to average over a combinatorial number
of equivalent assignments, and this ambiguity severely hinders
convergence. Latent tokens therefore need spatial positions as well. The natural
candidate, the ground-truth
vertex coordinates, cannot serve this purpose---they are exactly what
generation must produce and are unknown at inference. We therefore assign
each latent token a position from a deterministic Sobol point
set~\cite{sobol1967distribution} with the
same cardinality as the latent slots. During training, the real vertices,
represented by their continuous grid-space coordinates, are matched to the
Sobol candidates by an optimal-transport assignment minimizing squared
Euclidean cost. At inference, no ground truth is needed: the Sobol point set
itself provides the latent token positions.

\paragraph{Implementation details}
With the existence channel, the flow operates on $C{+}1 = 33$-channel
latent tokens. On the condition side,
image augmentation includes crop, outline, grayscale, blur, color jitter,
view sampling, and optional styled render packs. The flow is trained with
FSDP2 sharding, and batches are packed under a token cap rather than a
fixed sample count.

\section{Evaluation}
\label{sec:experiments}

We evaluate \methodname{} on a curated benchmark of 115 assets spanning objects,
architecture, creatures, and characters.
Each sample provides a reference photograph; the corresponding high-resolution geometry is
generated by Meshy~6 and uniformly decimated to a 100k-triangle ground truth.
Methods that accept a face budget are evaluated at roughly 4{,}000 faces ($\sim$2{,}000
vertices).
We compare eight pipelines in two tasks:
\textbf{high-poly retopology}, which must fit the high-resolution structure while producing
an artist-ready mesh, and \textbf{image-to-mesh generation}, which must recover appearance
from the photograph alone.
Diffusion baselines are Tripo~P1~\cite{wang2026nexus} and MeshFlow~\cite{li2026meshflowvae}; autoregressive
(AR) baselines are MeshAnything~V2~\cite{chen2024meshanythingv2}, BPT~\cite{weng2024bpt},
DeepMesh~\cite{zhao2025deepmesh}, MeshSilksong~\cite{songtopology}, and
FastMesh~\cite{kim2025fastmesh}.

\paragraph{Metrics}
Geometric fidelity uses bidirectional Chamfer Distance (CD) and Hausdorff Distance (HD)
between 10{,}000 uniformly sampled surface points on the prediction and a bbox-normalized
ground-truth mesh.
\textbf{Normal Consistency} (NC) averages the absolute dot product of bidirectionally
nearest-neighbor face normals on the same samples (higher is better).
Perceptual alignment uses Fr\'{e}chet Distance (FD)~\cite{heusel2017gans} on wireframe
renders against the input photographs, with both
Inception-v3~\cite{szegedy2016rethinking} and DINOv2
ViT-B/14~\cite{oquab2023dinov2} backbones.
Mesh usability reports mean non-manifold edge count and the fraction of triangles mergeable
by Blender's default \texttt{tris\_convert\_to\_quads}.
We also report median end-to-end time and success rate; a run counts as failed if it exceeds
\textbf{20 minutes}, raises a CUDA error, or produces no triangle faces.

\subsection{Position Encoding and Optimal Transport}
\label{sec:eval:pe_ot}

\begin{figure}[bh]
\centering
\includegraphics[width=\linewidth]{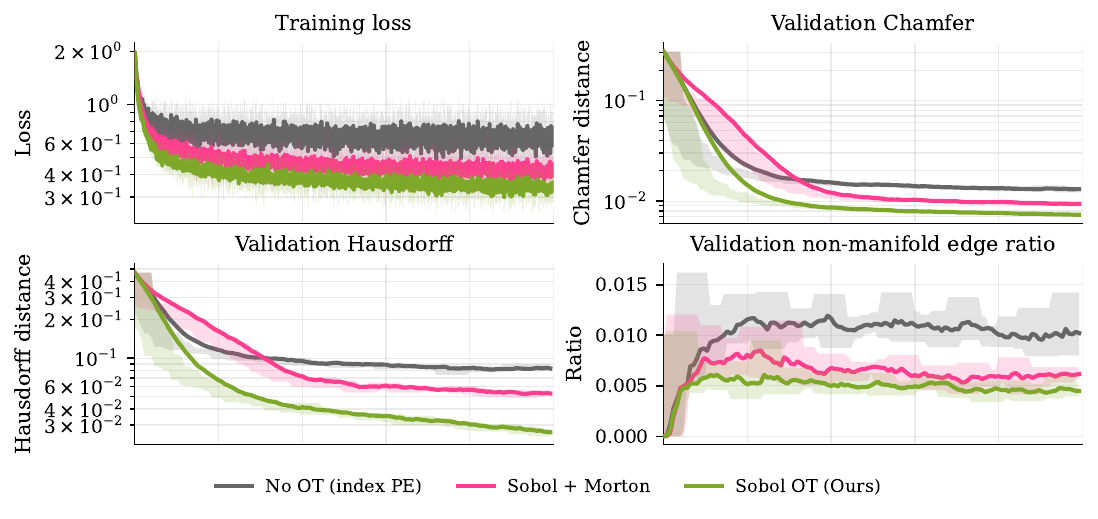}
\caption{\textbf{Position-encoding / OT ablation on the vertex-set mesh VAE.}
Solid lines show 10-tap EMA; shaded bands show the 10-step min--max envelope of raw logged
values. Sobol OT converges to lower validation Chamfer, Hausdorff, and non-manifold edge ratio
than index PE or Morton-order pairing.}
\label{fig:pe_ot_ablation}
\end{figure}

The vertex-set VAE processes an unordered latent token set, yet the transformer backbone
relies on 3D RoPE coordinates (Sec.~\ref{sec:method:pe}).
We ablate three position-encoding strategies on an otherwise identical VAE:
\textbf{No OT} binds RoPE to token index order;
\textbf{Sobol + Morton} uses the same Sobol candidates but skips the transport
assignment: vertices and candidates are each sorted in Morton order and paired
sequentially, which is cheaper but yields less accurate matches;
\textbf{Sobol OT} is our default, solving a minimum-cost assignment between
vertex coordinates and Sobol candidates.
Figure~\ref{fig:pe_ot_ablation} smooths
each series with a 10-tap exponential moving average and shades a per-step min--max envelope
over the same 10-step window.

\begin{table}[t]
\centering
\setlength{\tabcolsep}{9pt}
\renewcommand{\arraystretch}{1.15}
\begin{tabular}{lccc}
\toprule
Position encoding & Chamfer $\downarrow$ & Hausdorff $\downarrow$ & Non-manifold edge ratio $\downarrow$ \\
\midrule
No OT (index PE) & 0.0127 & 0.0783 & 0.0102 \\
Sobol + Morton & 0.0091 & 0.0534 & 0.0064 \\
\textbf{Sobol OT (Ours)} & \textbf{0.0070} & \textbf{0.0225} & \textbf{0.0043} \\
\bottomrule
\end{tabular}
\caption{\textbf{Position-encoding / optimal-transport ablation on the vertex-set mesh VAE.}
We compare three strategies for assigning 3D RoPE coordinates to unordered latent tokens
(Sec.~\ref{sec:method:pe}): no OT (\emph{index PE}), Sobol candidates paired
in Morton order, and full Sobol OT (\methodname{} default).
All runs share the same architecture and training recipe; metrics are final validation values.}
\label{tab:pe_ot_ablation}
\end{table}

\paragraph{Results}
Sobol OT improves every reported validation metric over both baselines
(Table~\ref{tab:pe_ot_ablation}).
Final Chamfer distance drops from 0.0127 (no OT) to \textbf{0.0070}, a 45\% reduction;
Hausdorff distance falls from 0.0783 to \textbf{0.0225}, more than $3\times$ lower.
Non-manifold edge ratio decreases from 0.0102 to \textbf{0.0043}.
Compared with Sobol + Morton, Sobol OT still yields 23\% lower Chamfer and 58\% lower
Hausdorff, indicating that the full transport assignment---not Sobol sampling alone---drives
most of the gain.
Training-loss curves show that Sobol OT also reaches a lower asymptotic loss than the
baselines, but the largest gaps appear on validation geometry and topology rather than on
optimization speed alone.
We therefore adopt Sobol OT as the default position-encoding strategy for both the VAE and
the second-stage latent flow.

\subsection{High-Poly Mesh Retopology}
\label{sec:eval:retopo}

In the retopology task, \methodname{} and MeshFlow consume both the reference image and the
100k high-resolution mesh and output a compact artist mesh aligned to the same ground truth.
AR methods receive the 100k mesh (or an equivalent MC-preprocessed input) and reconstruct an
artist mesh under the same CD / HD / NC protocol.
Tripo~P1 supports image-to-mesh generation only and is omitted from
Table~\ref{tab:retopo_alignment}.

\begin{table*}[bh]
\centering
\setlength{\tabcolsep}{5.5pt}
\renewcommand{\arraystretch}{1.12}
\footnotesize
\begin{tabular}{lccccccc}
\toprule
Method & CD $\downarrow$ & HD $\downarrow$ & NC $\uparrow$ &
Non-manifold $\downarrow$ & Tri-to-Quad $\uparrow$ &
Time (s) $\downarrow$ & Success $\uparrow$ \\
\midrule
\textbf{\methodname{}} & \textbf{0.020} & \textbf{0.044} & \textbf{0.860} &
\textbf{0.14} & 70.1\% & \textbf{3} & \textbf{100\%} \\
MeshFlow~\cite{li2026meshflowvae} & 0.319 & 0.534 & 0.595 &
304.5 & 45.2\% & 94 & \textbf{100\%} \\
\midrule
MeshAnything V2~\cite{chen2024meshanythingv2} & 0.037 & 0.151 & 0.833 &
69.6 & 75.0\% & 49 & \textbf{100\%} \\
BPT~\cite{weng2024bpt} & 0.061 & 0.136 & 0.797 &
39.0 & 60.9\% & 210 & 95.7\% \\
DeepMesh~\cite{zhao2025deepmesh} & 0.453 & 0.719 & 0.537 &
233.0 & \textbf{80.8\%} & 636 & 28.7\% \\
MeshSilksong~\cite{songtopology} & 0.309 & 0.525 & 0.629 &
19823.0 & 45.7\% & 1210 & 49.6\% \\
FastMesh~\cite{kim2025fastmesh} & 0.049 & 0.175 & 0.774 &
17234.7 & 17.4\% & 80 & \textbf{100\%} \\
\bottomrule
\end{tabular}
\caption{\textbf{High-poly mesh retopology results.}
Ground-truth geometry is a Meshy~6 high-resolution mesh decimated to 100k triangles.
\methodname{} and MeshFlow report retopologized artist meshes at $\sim$4{,}000 faces;
autoregressive baselines reconstruct an artist mesh from the 100k dense input.
Tripo~P1 provides image-to-mesh only and is excluded.
The Meshy~6 ground-truth row is a \emph{reference baseline only} and is excluded from best-metric ranking.
Normal Consistency (NC) measures bidirectional nearest-neighbor face-normal agreement
(10k surface samples; higher is better).
Non-manifold counts are means over successful outputs; high-face-count AR meshes are not
directly comparable in absolute count (see text).
Time is median seconds on successful samples; success rate counts outputs that finish
within 20 minutes without CUDA errors and contain at least one triangle.}
\label{tab:retopo_alignment}
\end{table*}

\paragraph{Geometric fidelity}
\methodname{} achieves the best overall geometric alignment: CD \textbf{0.020}, HD
\textbf{0.044}, and NC \textbf{0.860}.
MeshAnything~V2 ranks second on CD (0.037) because it conditions on the full dense mesh,
while FastMesh (0.049) and BPT (0.061) trail further behind.
Among diffusion methods with retopology, \methodname{} leads MeshFlow by a wide margin
(0.020 \vs{} 0.319 CD).
DeepMesh, MeshSilksong, and MeshFlow exhibit the weakest NC and CD scores.

\paragraph{Topology and robustness}
On retopologized \methodname{} outputs, mean non-manifold edges drop to \textbf{0.14} and
tri-to-quad mergeability reaches 70.1\%, reflecting the cleaner topology of the final artist
mesh rather than the raw image-to-mesh stage.
\methodname{}, MeshFlow, MeshAnything~V2, and FastMesh all reach \textbf{100\%} success within
the 20-minute budget; BPT succeeds on 95.7\% of assets (five large structures fail to
reconstruct).
DeepMesh completes only 28.7\% of runs under the robustness rule despite long wall-clock times
(median 636 s on logged successes), and MeshSilksong reaches 49.6\% because many outputs are
non-triangular point clouds.
Absolute non-manifold counts for FastMesh and MeshSilksong are inflated by very dense outputs
and should not be compared directly to low-face artist meshes.

\paragraph{Latency}
\methodname{} retopology runs in a median of \textbf{3 seconds} per asset,
more than an order of magnitude faster than MeshAnything~V2 (49 s), BPT (210 s), or MeshFlow
(94 s), while also delivering the strongest geometry.

\subsection{Image-to-Mesh Generation}
\label{sec:eval:i2t3d}

The image-to-mesh task evaluates whether a pipeline can reproduce the reference photograph
when only the image is available at inference time.
Table~\ref{tab:i2t3d_fd} reports FD against the input photos; the Meshy~6 row is included
only as a non-competitive reference baseline.
Native image-to-mesh diffusion methods are \methodname{}, Tripo~P1, and MeshFlow; AR baselines
are evaluated through a two-stage pipeline (image $\rightarrow$ dense mesh $\rightarrow$
artist mesh) and are included for cross-method comparison of rendered appearance.

\begin{table*}[bh]
\centering
\setlength{\tabcolsep}{7pt}
\renewcommand{\arraystretch}{1.12}
\footnotesize
\begin{tabular}{lcccc}
\toprule
Method & FD (Inception) $\downarrow$ & FD (DINOv2) $\downarrow$ &
Time (s) $\downarrow$ & Success $\uparrow$ \\
\midrule
Meshy~6 (reference) & 240.65 & 2021.68 & --- & --- \\
\midrule
\textbf{\methodname{}} & 255.77 & \textbf{2312.01} & \textbf{6} & \textbf{100\%} \\
Tripo~P1 & 255.76 & 2442.27 & 12 & \textbf{100\%} \\
MeshFlow~\cite{li2026meshflowvae} & \textbf{254.06} & 2577.00 & 94 & \textbf{100\%} \\
\midrule
MeshAnything V2~\cite{chen2024meshanythingv2} & 257.02 & 2499.05 & 49 & \textbf{100\%} \\
BPT~\cite{weng2024bpt} & 257.95 & 2457.99 & 210 & 95.7\% \\
DeepMesh~\cite{zhao2025deepmesh} & 281.68 & 2752.54 & 636 & 28.7\% \\
MeshSilksong~\cite{songtopology} & 301.09 & 2525.77 & 1210 & 49.6\% \\
FastMesh~\cite{kim2025fastmesh} & 268.21 & 2405.57 & 80 & \textbf{100\%} \\
\bottomrule
\end{tabular}
\caption{\textbf{Image-to-mesh generation results.}
Fr\'{e}chet Distance (FD) is computed on renders against the input reference
photographs under a fixed front-facing camera.
The Meshy~6 high-resolution mesh row is a \emph{reference baseline only} and is excluded
from best-FD ranking; bold values mark the best score among evaluated generation methods.
Diffusion methods (rows 2--4) generate meshes directly from a single image;
autoregressive methods first produce a 100k dense mesh from the image and then decode an
artist mesh, so their FD scores measure appearance after this two-stage pipeline rather
than native image conditioning.
Time is end-to-end median seconds for the evaluated pipeline; success rate uses the same
robustness rule as Table~\ref{tab:retopo_alignment}.}
\label{tab:i2t3d_fd}
\end{table*}

\paragraph{Perceptual alignment}
The Meshy~6 reference row (FD 240.65 / 2021.68) shows the empirical ceiling from
high-resolution mesh renders and is not treated as a competing method.
Among evaluated pipelines, MeshFlow achieves the lowest Inception FD (254.06), while
\methodname{} leads on DINOv2 (2312.01 \vs{} 2442.27 for Tripo~P1 and 2577.00 for
MeshFlow), indicating stronger semantic and structural alignment with the source images.
DeepMesh and MeshSilksong deviate most from the photograph distribution; FastMesh sits between
the stronger AR and diffusion methods on DINOv2 (2405.57).

\paragraph{Latency and reliability}
\methodname{} completes end-to-end image-to-mesh generation in a median of \textbf{6 s} with
\textbf{100\%} success, faster than Tripo~P1 (12 s) and substantially faster than every AR
baseline.
MeshFlow matches \methodname{} on reliability but requires 94 s per asset.
DeepMesh and MeshSilksong combine low success rates (28.7\% and 49.6\%) with median runtimes
exceeding 10 minutes, making them unsuitable for interactive asset creation despite occasional
strong tri-to-quad ratios on successful DeepMesh outputs.

\section{Conclusion}
\label{sec:conclusion}

We presented \methodname{}, a fast mesh generation framework for compact
mesh assets. Instead of framing the contribution as direct mesh generation
alone, \methodname{} emphasizes the practical capabilities enabled by joint
vertex-connectivity modeling. The system generates meshes within ten seconds,
exposes strong face-count control by sampling a requested number of vertex
slots before decoding, and uses the triangle-mesh Euler relation
$F \approx 2V$ to make the resulting face budget predictable.

By directly encoding the connectivity observed in training meshes, the generated
vertex-edge graph naturally decomposes into connected components. This lets the
model produce multi-component assets without an additional component-wise
generation, splitting, or stitching stage. Future work will further improve
topological robustness on highly irregular source meshes, extend the framework
to richer material and part-level controls, and study scene-level generation
where multiple compact meshes must be produced under a shared global budget.

\paragraph{Acknowledgments} We thank Hao Jiang and Jianqiao Gong for
insightful discussions and valuable suggestions. The ``Lighthouse'' photo was
taken by Walter Rademacher, from
\href{https://commons.wikimedia.org/wiki/File:Aerial_photograph_60D_2012_05_13_8760_DxO_retusche.jpg}{Wikimedia
Commons},
licensed under CC BY-SA 4.0.



\renewcommand{\refname}{{\color{MeshyLimeDark}References}}






\renewcommand{\refname}{{\color{MeshyLimeDark}References}}

\bibliographystyle{plainnat}
\bibliography{references}

\clearpage
\end{document}